# Local Group dwarf galaxies as dark matter probes

**Authors:** G. Battaglia[1,2], J.M. Arroyo Polonio[1,2], R. Pascale[3], M. Benito[1,2], R. Leaman[4], G. Thomas[1,2]

**Affiliations:**

[1] Instituto de Astrofísica de Canarias, Calle Vía Láctea s/n, 38206 La Laguna, Santa Cruz de Tenerife, Spain

[2] Universidad de La Laguna, Avda. Astrofísico Francisco Sánchez 38205 La Laguna, Santa Cruz de Tenerife, Spain

[3] INAF – Osservatorio di Astrofisica e Scienza dello Spazio di Bologna, Via Piero Gobetti 93/3, 40129 Bologna, Italy

[4] Department of Astrophysics, University of Vienna, Türkenschanzstrasse 17, A-1180 Wien, Austria

## Summary

Unveiling the fundamental nature of non-baryonic dark matter (DM) has profound implications for our understanding of the Universe and of the physical laws that govern it. Its manifestation as an additional source of matter necessary to explain astrophysical and cosmological observations indicates either a breakdown of General Relativity or that the current Standard Model of Particle Physics is incomplete.

In the standard Cold DM (CDM) paradigm, DM consists of collisionless non-relativistic particles with negligible non-gravitational interactions. This simple hypothesis is very successful on large and intermediate scales, but faces challenges on small galactic scales.

Local Group (LG) dwarf galaxies can play a fundamental role to elucidate whether these challenges stem from poorly understood fundamental baryonic processes or instead indicate that alternative DM scenarios need to be considered. In particular, a systematic determination of their DM halo properties as a function of stellar mass and star formation histories (SFH) will provide crucial observational benchmarks for models to deal with the trickiest issue that prevents us from advancing in our understanding of DM nature, i.e. the impact of baryonic processes in altering the properties of the inner regions of DM haloes.

Such systematic study would require assembling accurate l.o.s. velocities (and metallicities) for several thousands of stars per galaxy, for an heterogeneous sample of target galaxies, spanning order of magnitudes in stellar mass and covering distances from about 100 kpc to more than 1 Mpc. This calls for both multi-objects spectrographs on 12m-class telescopes with fov of a few deg² and a multi-plex power in the several 1000s with the capability of providing dense sampling of the innermost regions, as well as for wide-area multi-objects spectrographs with fov of several arcmin² on 30-40m class telescopes.

## Scientific Context

Determining the fundamental nature of DM has been identified as a priority for the future of astrophysics, cosmology and particle physics, driving extensive observational and theoretical efforts worldwide.

The simple CDM hypothesis, successful on large and intermediate scales, faces challenges on small galactic scales. In fact, observations of stellar and gas kinematics in low surface brightness dwarf galaxies suggest that some may occupy DM haloes less dense than if they would follow the DM halo density profiles and sub-halo mass functions predicted in CDM (e.g. [1], [2]). These are commonly known as the *cusp/core and too-big-to-fail problems* (see detailed discussion in [3]).

Alternative DM models (e.g. self-interacting: [4], [5]; fuzzy dark matter: [6], [7]) may offer potential solutions to these problems as they can produce DM cores at those scales. Nonetheless, also baryonic processes can be responsible of modifying the inner structure of DM haloes. In particular, rapid and violent variations of the gravitational potential due to gas expulsion as a consequence of supernovae feedback has been identified as a key physical mechanism for decreasing the inner DM halo density and even leading to the formation of constant density DM cores ([8], [9], [10]). This same mechanism is also relevant when interpreting findings on the DM halo flattening, as it can lead to 'sphericization' of the halo ([11]) with the mechanism efficient at both low ($M\star<10^8$ $M_\odot$; [12], [13]) and high mass ($M\star>10^8$ $M_\odot$, ; [14], [15]).

Within a given DM framework, there is no general agreement in the predictions from numerical simulations on how redistribution of DM due to baryonic effects should occur or whether it should occur at all (e.g. [16], [17]; for a review see [18]). This is mainly due to difficulties in determining the efficiency of stellar feedback observationally and by the need to resort to sub-grid treatment in simulations. In those CDM simulations that form DM cores, in general the efficiency of the process is found to depend on the stellar-to-halo mass ratio $M\star/M_{halo}$ (i.e. on the ratio of total energy injected versus total gravitational potential), with a peak around $10^{-3} < M\star/M_{halo} < 10^{-2}$ ([19], [20]). However, several aspects are debated, such as the expected core sizes, the minimum mass to form a core and its relation to other parameters. For example, recent simulations suggest that the duration of star formation plays a strong role in the process of core creation ([21]). Given the uncertain theoretical landscape, *observational benchmarks of how the DM halo properties vary as a function of the main parameters involved in stellar-feedback driven DM halo alterations ($M\star$ and SF duration, over the relevant regime of stellar masses, $M\star < 10^8$ $M_\odot$) are absolutely vital for this long standing issue to be solved, pushing down to the regime in which stellar feedback is thought to become ineffective and DM halo properties should be directly indicative of the nature of DM ($M\star < 10^4$ -$10^5$ $M_\odot$)*.

The wealth of dwarf galaxies with stellar masses $M\star < 10^8$ $M_\odot$ (down to 200 $M\star$, [22]) and a variety of SFHs (e.g. [23]) found in the LG and immediate vicinity provide the exciting opportunity of mapping DM halo properties as a function of those parameters.

Stars are often the only accessible kinematic tracer for LG dwarf galaxies. Most low mass LG dwarf galaxies are either devoid of neutral gas and have an HI component unsuitable for detailed dynamical studies. Nonetheless, in those cases when both stellar and HI kinematics can be simultaneously used, this leads to an important gain in the precision of the recovered DM halo profile and opens the tantalizing possibility of placing constraints on the DM halo shape (e.g. [24]).

Samples of the order of 1000 *3D* velocities with precisions of the order of 2 km/s *over a wide-area* would provide the ideal data-set for precise determinations of the dark matter halo density profiles (e.g. [25]). In the next decade there will be the unprecedented opportunity to assemble such data for a few of the brighter Milky Way satellite galaxies, thanks to tangential velocities from proper motions obtained with Nancy Roman, Gaia DR5, along with combinations of Euclid, JWST data at various epochs (e.g. [26], [27] for Gaia-HST and HST-HST studies). **Nonetheless, stellar spectroscopy is and will remain the absolute key,** as radial velocities of individual stars will remain the most precise, and often the only accessible, probe of the internal kinematic properties of most LG dwarf galaxies. It has been shown that the stellar component of several of the brighter MW satellites can be described as the super-position of two or more stellar populations of different spatial distribution, mean metallicity and kinematic properties (e.g. [28], [29], [30]) and that these components can be modeled together to provide tighter constraints on these type of spheroidal, non-rotating galaxies (e.g. [31], [32], [33]).

L.o.s. velocities with precision better than 1-2 km/s and metallicities with precision better than 0.05-0.1dex are needed both to unveil such multiple chemo-kinematic populations across the sample of LG dwarf galaxies, safely assigning stars to these components and using them for DM inferences. Our own dynamical modeling tests on mock catalogues, where we have applied the most detailed dynamical modeling methodology available ([33]), show that an increase from 500 to 5000 stars, with a velocity precision better than 1-2 km/s and [Fe/H] to 0.05-0.1dex, leads to twice better precision in the recovered DM halo density at all radii and to unbiased recoveries. Importantly our analysis (in prep.) shows that the spatial distribution of the target stars is a crucial factor, with the highest precision in the DM density recovery being achieved in the regions most densely sampled. This calls for a dense sampling from the innermost to the outermost regions, with the former being most informative on the nature of DM vs baryonic processes, and the latter on the larger scale mass content.

## Facilities

In order to assemble a large sample of dwarf galaxies spanning 4 orders of magnitude in stellar mass ($10^4$ $M_\odot$ < M⋆ < $10^8$ $M_\odot$) and probing different SFHs at given stellar masses, it is necessary to have access not only to the MW system of satellite galaxies but also the M31 system and isolated LG dwarf galaxies, i.e. spanning a distance range from about 100 kpc to more than 1 Mpc.

The resulting heterogeneity in the stellar mass, surface brightness and distance calls for a multi-facility approach. For example, for an old and metal-poor stellar system (simulated with IAC-Basti) of $10^6$ $L_\odot$ the 5000 brightest stars would be found down to G ~ 21.5 at 100kpc and G ~ 26.5 at 1Mpc; while assembling ~1000 targets in a $10^4$ $L_\odot$ system at 100 kpc would require reaching as faint as G~25. **Multi-objects spectrographs on 12m-class telescopes with fov of a few deg² and a multi-plex power of several 1000s, accompanied by the capability of providing dense sampling of the innermost regions** (for example with massive integral field units), will be essential to gather in an efficient manner the spectroscopic data that will have the statistics, velocity precision and spatial distribution needed to perform accurate and precise determinations of the DM halo properties *of the brighter dwarf galaxies, across the whole distance range*, through the application of the most sophisticated and constraining dynamical modeling methodologies (e.g. [33], [34]); if such facilities could reach down to G~23, they would bring dynamical modeling of the bright MW satellite galaxies into a completely new regime, with ~$10^4$ stars accessible for a $10^6$ $L_\odot$ system at 100 kpc. In addition, **wide-field spectrographs with fov of arcmin² on 30m-class telescopes will be the key facilities to gather the needed spectroscopic data-sets for the galaxies at the faintest end in the surrounding of the MW, M⋆ < $10^4$ -$10^5$ $M_\odot$, and for systems with M⋆ around $10^6$ $M_\odot$ found in the M31 system and in isolation.**